\documentclass[prb,aps,twocolumn,amsmath,amssymb,floatfix,superscriptaddress]{revtex4}
\usepackage[dvips]{graphics}
\usepackage{color}
\definecolor{dred}{rgb}{0,0,0.6}
\usepackage{amsmath}

\begin{document}

\title{Effect of spin-orbit interaction on circular current: Pure spin current phenomena within a ring
conductor}

\author{Moumita Patra}

\affiliation{Department of Physics, Indian Institute of Science Education and Research, Pune 411008, India}

\begin{abstract}

A net circulating current may appear within a quantum ring under finite bias. We study
the characteristic features of the circular current in the presence of Rashba spin-orbit
interaction (RSOI). Both charge and spin currents appear within the ring. Whereas when the
ring is symmetrically connected to the external leads, we can get a pure charge current
at non-zero Fermi-energy. On the other hand, for asymmetric ring-to-leads configuration,
at zero Fermi-energy, the spin current vanishes but a pure charge current flows within the ring.
Tuning RSOI, we demonstrate a way to control the pure spin
current externally. This new perspective of the generation of the pure spin circular
current can open a new basis for the highly efficient, low energy cost spintronic
devices.

\end{abstract}

\maketitle

\section{Introduction}

In the context of quantum transport, we generally
focus on the overall conduction properties of a junction. But when the bridging conductor
contains a loop structure, there is a possibility to induce a circular current within the loop.
Circular current may behave very differently and may have a very large magnitude
compared to the overall drain current. The circular current may appear within a quantum loop
under several circumstances. In the early 80's B\"{u}ttiker et al.~\cite{cir1} first
proposed theoretically that a small conducting ring carries a net circulating charge current,
commonly known as persistent current in the presence of the magnetic field. Followed by this, there were
lots of theoretical as well as experimental propositions~\cite{cir2,cir3,cir4,cir5,cir6,cir7}
in this direction. Using phase-locked infra-red laser pulses circular current has been
generated in an isolated quantum ring\cite{cir8}. Several other theoretical works have
indicated the possibility to excite such loop currents by using external
radiation~\cite{cir9}, shaped photon pulses~\cite{cir10,cir11}, circularly
polarized light~\cite{cir11a}, twisted light~\cite{cir11b}, etc. Circular current can be
also induced in quantum rings driven by an external voltage~\cite{cir11c,cir11d,cir11e,cir11f,cir11g,cir11h}.
For example, S. Nakanishi and M.
Tsukada~\cite{cir12} have predicted the existence of a quantum internal current through the
$\mbox{C}_60$ molecular bridge. Large loop currents circulating around the zigzag and
chiral carbon nanotubes have been observed by N. Tsuji et.al.~\cite{cir13}. The circular
currents due to different driving forces are closely related in nature.

Though the idea of bias induced circular current is so far limited to theoretical
computations, but it involves various important factors in the context of
quantum transport. Such as, it gives the measurement of current through the
individual section of a complicated quantum loop system consists of multiple
pathways. Depending on the voltage bias, circular current may rise to a very
high value compared to the overall drain current ($\sim10^3$ times larger) at the
outgoing leads. This giant circular
current induces a large magnetic field (in some cases it may even reach to few
millitesla or even Tesla) at the center of the ring, which is very important
\begin{figure}[ht]
{\centering \resizebox*{8cm}{3.2cm}{\includegraphics{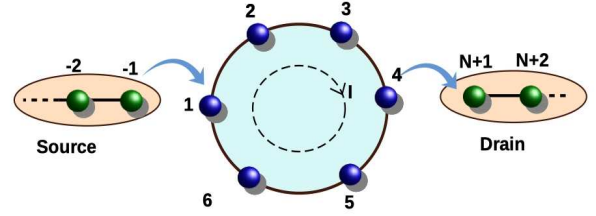}}\par}
\caption{(Color online). Schematic-representation of the Tight-Binding quantum ring, attached
with two semi-infinite electrodes.}
\label{model}
\end{figure}
in the context of local spin regulation and several other electronic and
spintronic applications like storage of data, logic functions, spin switching,
spin-selective electron transmission, spin-based quantum computations,
etc~\cite{app1,app2,app3,app4,app5}.

In a recent work~\cite{cir11h}, the idea of spin circular currents
has been addressed, where the spin components have been defined by the conservation
law between the bond current and
transport current in a one-dimensional quantum chain. With this formulation, 
here we make an in-depth analysis on the effect of the Rashba spin-orbit
interaction (RSOI) on the bias induced-circular current. RSOI is originated due to the
structure inversion asymmetry caused by the inversion asymmetry of the confining
potential~\cite{Rashba0}. It~\cite{Rashba1,Rashba2} is an electrically tunable spin-orbit
interaction~\cite{RashbaTune}.

Generation of pure spin current is the ultimate requirement for the spintronic devices,
which have evolved from exploiting spin-polarized current to pure spin current.
It helps in gaining speed, miniaturization, and high energy
efficiency~\cite{PureSpin1,PureSpin2} as in this case only electron-spin
carries the information. The energy dissipation due to Joule heating, which is the
main source of the power dissipation in conventional electronic devices, can be
completely suppressed here. It also allows to have spin–orbit torque, different from
the spin-transfer torque, which can switch ferromagnetic free layers to design high-density
memory devices~\cite{ps1}. The spin Hall effect~\cite{ps2}, spin pumping~\cite{ps3}, ferromagnetic and
anti-ferromagnetic metals and insulators are the few ways to generate pure spin current.
Here we propose a new idea to generate pure spin current using RSOI. The model is composed of a
quantum wire attached to two external baths as shown in Fig.~\ref{model}. The entire system is
non-magnetic and metallic. RSOI is considered at the bridging ring. Under the symmetric ring-to-lead
configuration (when the length of the upper arm of the ring is equal to the length of
the lower one) the system has two-fold degeneracy along with the spin-degeneracy.
In this situation, the currents at the two arms of the ring are equal and opposite to each
other, resulting in a zero charge circular current. In the presence of RSOI,
when unpolarized electrons are injected, it becomes polarized within the ring in
such a way that the charge current becomes zero. Hence a pure spin current is
generated within the ring. Here the up spin moves to the opposite direction in the down spin.
Though the outgoing drain current always remains unpolarized for symmetric as well as asymmetric
configurations (when the arm lengths of the ring are unequal). The spin current density
is anti-symmetric around incident energy $E$ equals to 0. Therefore we need to set
a non-zero Fermi energy to get pure spin circular current.

The conversion of the pure spin current to the pure charge current is also
possible here. In an asymmetric junction, if we set the Fermi-energy at zero,
the spin current vanishes, resulting in a pure charge current. We find that
the spin current is very robust against the connection positions of the
electrodes to the ring, unlike the charge current. To make the spin-based
quantum computers and other spintronic devices, proper spin regulation
is highly important. Tuning the strength of RSOI, we propose a suitable
way to control the pure spin current externally. Based on the tight-binding
(TB) framework we compute the circular current using
wave-guide formalism~\cite{cir11g,cir11h,wave1,wave2,wave3}. With this approach, one
can find current carried by each section of the ring. Circular current may
decrease with voltage (showing negative differential resistance, NMR) contrary
to the overall drain current which increases with voltage.

The arrangement of the remaining part is as follows. In sec. II we
thoroughly discussed the methodology to calculate the current in the
presence of SOI. In sec. III, we illustrate all the
essential results, and finally, we summarize our findings in sec. IV.

\section{The model and theory}

\subsection{Hamiltonians}

The Hamiltonian $\mbox{\boldmath $H$}$ for the entire system (shown in Fig.~\ref{model})
can be written as the sum of the Hamiltonians for the ring $\mbox{\boldmath $H_R$}$, the
electrodes (namely, source S and drain D) $\mbox{\boldmath $H_{(S/D)}$}$, and the tunneling
between the ring and electrodes $\mbox{\boldmath $H_T$}$. Therefore,
\begin{eqnarray}
\mbox{\boldmath $H$} & = & \mbox{\boldmath $H_R$} + \mbox{\boldmath $H_{(S/D)}$}
+ \mbox{\boldmath $H_T$}.
\label{h0}
\end{eqnarray}
$\mbox{\boldmath $H_R$}$ represents the Hamiltonian for a one-dimensional quantum
ring with spin-orbit interaction (SOI), having the TB~\cite{tb1,tb2} form:
\begin{eqnarray}
\mbox{\boldmath $H_R$} & = &\sum\limits_{n} \mbox{\boldmath $c$}_n^\dagger
\mbox{\boldmath $\epsilon$}_n \mbox{\boldmath $c$}_n 
+ \sum\limits_{n}\left( \mbox{\boldmath $c$}_{n+1}^\dagger\mbox{\boldmath $t$}
\mbox{\boldmath $c$}_n + \mbox{\boldmath $c$}_{n}^\dagger\mbox{\boldmath $t$}
\mbox{\boldmath $c$}_{n+1} \right) \nonumber \\
&  & -\sum\limits_{n}\left(\mbox{\boldmath $c$}_{n+1}^\dagger\left(\imath \mbox{\boldmath $\sigma$}_x\right)
\mbox{\boldmath $\alpha$} \cos \phi_{n,n+1} \mbox{\boldmath $c$}_n + h.c.\right)\nonumber \\
& & -\sum\limits_{n}\left(\mbox{\boldmath $c$}_{n+1}^\dagger\left(\imath \mbox{\boldmath $\sigma$}_y\right)
\mbox{\boldmath $\alpha$} \sin \phi_{n,n+1} \mbox{\boldmath $c$}_n + h.c.\right).
\label{hamil}
\end{eqnarray}
\vskip 0.01cm
\noindent
$n$ is the site-index runs from $1$ to $N$, where $N$ is the number of sites in the ring.
The other factors are: 
\begin{center}
$\mbox{\boldmath $\epsilon$}_n=\left(\begin{array}{cc}
    \epsilon_{n,\uparrow} & 0 \\ 
    0 & \epsilon_{n,\downarrow}
\end{array}\right)$, $\mbox{\boldmath $t$}=\left(\begin{array}{cc}
    t & 0 \\ 
    0 & t
\end{array}\right)$, \\
$\mbox{\boldmath $c$}_n=\left(\begin{array}{cc}
    c_{n,\uparrow} \\ c_{n,\downarrow}
\end{array}\right)$, 
$\mbox{\boldmath $\alpha$}=\left(\begin{array}{cc}
    \alpha & 0 \\ 
    0 & \alpha
\end{array}\right).$
\end{center}
\noindent
$\epsilon_{n,\uparrow(\downarrow)}$ represents the on-site potential of an
up (down) spin  electron. We consider $\epsilon_{n,\uparrow}=\epsilon_{n,\downarrow}=\epsilon_n$
for the sake of simplicity. $\alpha$ is the Rashba spin-orbit coupling strength.
$\phi_{n,n+1}=\left(\phi_n+\phi_{n+1}\right)/2,$ with $\phi_n=2\pi(n-1)/N$.
$\mbox{\boldmath $\sigma$}_i$'s ($i=x$, $y$, $z$) are the Pauli spin 
matrices in $\mbox{\boldmath $\sigma$}_z$ diagonal representation. $\imath =\sqrt{-1}$.

The Hamiltonian $\mbox{\boldmath $H_{(S/D)}$}$, representing
the electrodes, characterized by the onsite potential $\mbox{\boldmath $\epsilon$}_0$
and the nearest neighbor hopping integral $\mbox{\boldmath $t$}_0$ has the form:
\begin{eqnarray}
\mbox{\boldmath $H_{S/D}$} & = & \sum\limits_{n\leq-1} \mbox{\boldmath $a$}_n^\dagger
\mbox{\boldmath $\epsilon$}_0 \mbox{\boldmath $a$}_n
+ \sum\limits_{n\leq-1}\left( \mbox{\boldmath $a$}_{n+1}^\dagger\mbox{\boldmath $t_0$}
\mbox{\boldmath $a$}_n + \mbox{\boldmath $a$}_{n}^\dagger\mbox{\boldmath $t_0$}
\mbox{\boldmath $a$}_{n+1} \right)\nonumber \\
& + & \sum\limits_{n\geq N+1} \mbox{\boldmath $b$}_n^\dagger
\mbox{\boldmath $\epsilon$}_0 \mbox{\boldmath $b$}_n
+ \sum\limits_{n\geq N+1}\left( \mbox{\boldmath $b$}_{n+1}^\dagger\mbox{\boldmath $t_0$}
\mbox{\boldmath $b$}_n + \mbox{\boldmath $b$}_{n}^\dagger\mbox{\boldmath $t_0$}
\mbox{\boldmath $b$}_{n+1} \right).\nonumber \\
& &
\label{sd}
\end{eqnarray}
\noindent
$\mbox{\boldmath $a$}_n (\mbox{\boldmath $b$}_n)$ and $\mbox{\boldmath $a$}_n^{\dagger}
(\mbox{\boldmath $b$}_n^{\dagger})$ are the annihilation and creation
operators, respectively of the source (drain).

$\mbox{\boldmath $H_T$}$ describes the coupling of
the ring with S and D, and it is also expressed in the usual TB form.

\subsection{Circular current density}

We evaluate the spin-dependent circular current density within the ring adopting wave-guide
theory, where we solve the Schr\"{o}dinger equation
\begin{eqnarray}
\mbox{\boldmath $H$}|\psi\rangle & = & \mbox{\boldmath $E$}\bf{I}|\psi\rangle.
\label{sch}
\end{eqnarray}
\noindent
$\mathbf{I}$ is the ($2\times2$) identity matrix. The wave function $|\psi\rangle$,
representing the entire system has the form:
\begin{equation}
|\psi\rangle =\left[\sum\limits_{n \le -1}\mbox{\boldmath $A$}_n
\mbox{\boldmath $a$}_n^{\dagger} + \sum\limits_{n \ge 1}
\mbox{\boldmath $B$}_n\mbox{\boldmath $b$}_n^{\dagger}
+ \sum\limits_{i=1}\mbox{\boldmath $C$}_i\mbox{\boldmath
$c$}_{i}^{\dagger}\right]|0\rangle.
\label{wave}
\end{equation}
The coefficients $\mbox{\boldmath $A$}_n =\left(\begin{array}{cc}
    A_{n,\sigma\sigma'} \\
    A_{n,\sigma\sigma'}
\end{array}\right)\,$, $\mbox{\boldmath $B$}_n=\left(\begin{array}{cc}
    B_{n,\sigma\sigma'} \\
    B_{n,\sigma\sigma'}
\end{array}\right)\,$, and $\mbox{\boldmath $C$}_i=\left(\begin{array}{cc}
    C_{i,\sigma\sigma'} \\
    C_{i,\sigma\sigma'}
\end{array}\right)\,$
correspond to the amplitudes for an electron at the $n$-th site of the source, drain,
and $i$-th site of the ring, respectively.
From Eq.~(\ref{sch}) we get a set of coupled equations as:
{
\allowdisplaybreaks
\begin{widetext}
{\begin{center}
\begin{eqnarray}
\left[\left(\begin{array}{cc}
        E & 0 \\
        0 & E
\end{array}\right) - \left(\begin{array}{cc}
    \epsilon_0 & 0 \\
    0 & \epsilon_0
\end{array}\right)\right]\left(\begin{array}{cc}
        A_{n,\sigma\sigma'} \\
        A_{n,\sigma\sigma'}
\end{array}\right)  & = &\left(\begin{array}{cc}
    t_0 & 0 \\
    0 & t_0
\end{array}\right) \left(\begin{array}{cc}
        A_{n+1,\sigma\sigma'}\\
        A_{n+1,\sigma\sigma}
    \end{array}\right) + \left(\begin{array}{cc}
   t_{0}  & 0 \\
    0 & t_{0}
\end{array}\right) \left(\begin{array}{cc}
   A_{n-1,\sigma\sigma'}  \\
   A_{n-1,\sigma\sigma'}
\end{array}\right),~~~~n\leq-2, \nonumber
\end{eqnarray}
\end{center}
\begin{center}
\begin{eqnarray}
\left[\left(\begin{array}{cc}
        E & 0 \\
        0 & E
\end{array}\right) - \left(\begin{array}{cc}
    \epsilon_0 & 0 \\
    0 & \epsilon_0
\end{array}\right)\right]\left(\begin{array}{cc}
        A_{-1,\sigma\sigma'} \\
        A_{-1,\sigma\sigma'}
\end{array}\right) &  =  &\left(\begin{array}{cc}
    t_0 & 0 \\
    0 & t_0
\end{array}\right) \left(\begin{array}{cc}
         A_{-2,\sigma\sigma'} \\
         A_{-2,\sigma\sigma'}
    \end{array}\right) + \left(\begin{array}{cc}
   t_{S}  & 0 \\
    0 & t_{S}
\end{array}\right) \left(\begin{array}{cc}
   C_{N_S,\sigma\sigma'} \\
   C_{N_S,\sigma\sigma'}
\end{array}\right),\nonumber
\end{eqnarray}
\end{center}
\begin{center}
\begin{eqnarray}
\left[\left(\begin{array}{cc}
        E & 0 \\
        0 & E
\end{array}\right) - \left(\begin{array}{cc}
    \epsilon_0 & 0 \\
    0 & \epsilon_0
\end{array}\right)\right]\left(\begin{array}{cc}
        B_{n,\sigma\sigma'} \\
        B_{n,\sigma\sigma'}
\end{array}\right) & = & \left(\begin{array}{cc}
    t_0 & 0 \\
    0 & t_0
\end{array}\right) \left(\begin{array}{cc}
        B_{n+1,\sigma\sigma'}\\
        B_{n+1,\sigma\sigma}
    \end{array}\right) + \left(\begin{array}{cc}
   t_{0}  & 0 \\
    0 & t_{0}
\end{array}\right) \left(\begin{array}{cc}
   B_{n-1,\sigma\sigma'}  \\
   B_{n-1,\sigma\sigma'}
\end{array}\right),~~~~n\geq2 \nonumber 
\end{eqnarray}
\end{center}
\begin{center}
\begin{eqnarray}
\left[\left(\begin{array}{cc}
        E & 0 \\
        0 & E
\end{array}\right) - \left(\begin{array}{cc}
    \epsilon_0 & 0 \\
    0 & \epsilon_0
\end{array}\right)\right]\left(\begin{array}{cc}
        B_{1,\sigma\sigma'} \\
        B_{1,\sigma\sigma'}
\end{array}\right)  & = & \left(\begin{array}{cc}
    t_0 & 0 \\
    0 & t_0
\end{array}\right) \left(\begin{array}{cc}
         B_{2,\sigma\sigma'} \\
         B_{2,\sigma\sigma'}
    \end{array}\right) + \left(\begin{array}{cc}
   t_{D}  & 0 \\
    0 & t_{D}
\end{array}\right) \left(\begin{array}{cc}
   C_{N_D,\sigma\sigma'}\\
    C_{N_D,\sigma\sigma'}
\end{array}\right),\nonumber
\end{eqnarray}
\end{center}
\begin{center}
\begin{eqnarray}
\left[\left(\begin{array}{cc}
    E & 0 \\
    0 & E
\end{array}\right) - \left(\begin{array}{cc}
    \epsilon_i & 0 \\
    0 & \epsilon_i
\end{array}\right)\right] \left(\begin{array}{cc}
   C_{i,\sigma\sigma} \\
    C_{i,\sigma\sigma'}
\end{array}\right) & = & \left(\begin{array}{cc}
    t & -\imath\alpha e^{-\imath\varphi_{i,i+1}} \\
    -\imath\alpha e^{-\imath\varphi_{i,i+1}} & t
\end{array}\right)\left(\begin{array}{cc}
     C_{i+1,\sigma\sigma} \\
        C_{i+1,\sigma\sigma}
    \end{array}\right) \nonumber \\
& + & \left(\begin{array}{cc}
    t & -\imath\alpha e^{-\imath\varphi_{i,i-1}} \\
    -\imath\alpha e^{-\imath\varphi_{i,i-1}} & t
\end{array}\right) \left(\begin{array}{cc}
   C_{i-1,\sigma\sigma'} \\
    C_{i-1,\sigma\sigma'}
\end{array}\right) \nonumber \\
& + & \left(\begin{array}{cc}
 t_S & 0 \\
    0 & t_S
\end{array}\right) \left(\begin{array}{cc}
   C_{N_S,\sigma\sigma'}  & 0 \\
    0 & C_{N_S,\sigma\sigma'}
\end{array}\right)\delta_{i,N_S} \nonumber \\
& + & \left(\begin{array}{cc}
 t_D & 0 \\
    0 & t_D
\end{array}\right) \left(\begin{array}{cc}
   C_{N_D,\sigma\sigma'}\\
    C_{N_D,\sigma\sigma'}
\end{array}\right)\delta_{i,N_D},~~~~1\leq i \leq N.
\label{eqn}
\end{eqnarray}
\end{center}}
\end{widetext}}
\noindent
$\sigma$ represents up and down spins and similarly $\sigma'$ also. $\mbox{$t_{S}$}$ and $\mbox{$t_{S}$}$
are the couplings between the source and the
drain to the $N_S$-th and $N_D$-th sites of the ring, respectively. 

\vskip 0.3cm
Depending upon the nature of incident electrons, now we consider two different situations.
\vskip 0.2cm
(i) {\em Up spin incidence from the source lead:}
\vskip 0.2cm
In this case, we consider that an up spin electron incidents as a plane wave with
unit amplitude, having the form:
\vskip 0.2cm
\noindent
$\mbox{\boldmath $A$}_n=\left(\begin{array}{cc}
    e^{ik(n+1)a} + r_{\uparrow\uparrow}e^{-ik(n+1)a} \\
    r_{\uparrow\downarrow}e^{-ik(n+1)a}
    \end{array}\right)$ and \\
$\mbox{\boldmath $B$}_n=\left(\begin{array}{cc}
    \tau_{\uparrow\uparrow}e^{ikna} \\
    \tau_{\uparrow\downarrow}e^{ikna}
    \end{array}\right)\,$,
\vskip 0.2cm
\noindent
where $a$ being the lattice spacing and $k$ is the wave vector associated with
the energy $E$. $\tau_{\uparrow\uparrow}$ ($\tau_{\uparrow\downarrow}$) and
$r_{\uparrow\uparrow}$ ($r_{\uparrow\downarrow}$) are the
transmission and reflection amplitudes of an up spin, transmitted, and reflected
as up (down) spin, respectively.

Putting the expression of $\mathbf{A}_n$ and $\mathbf{B}_n$ in
Eq.~(\ref{eqn}), we solve the wave amplitudes $C_{i,\uparrow\sigma}$s
and the transmission amplitudes $t_{\uparrow\sigma}$, $\sigma = \uparrow, \downarrow$
for a particular energy associated with wave vector $k\,$. We finally get the
spin-dependent transmission probability and the bond current density between
the sites $i$ and $i + 1$ of the ring as:
\begin{eqnarray}
T_{\uparrow\sigma} & = & |\tau_{\uparrow\sigma}|^2
\label{tran1}
\end{eqnarray}
and
\begin{eqnarray}
J_{i\rightarrow i+1\uparrow\sigma} & = & \frac{\sqrt{t^2+\alpha^2}\mbox{Im}
\left[\,C_{i,\uparrow\sigma}^*C_{i+1,\uparrow\sigma} \right]}{(1/2)t_0\sin(ka)},
\sigma\rightarrow\uparrow,\downarrow, \nonumber \\
\label{j1}
\end{eqnarray}
respectively.

\vskip 0.2cm
(ii) {\em Down spin incidence from the source lead :}
\vskip 0.2cm
For this case, down spin incidents with unit amplitudes, where $\mathbf{A}_n$ and
$\mathbf{B}_n$ look like:
\vskip 0.2cm
\noindent
$\mbox{\boldmath $A_n$}=\left(\begin{array}{cc}
    r_{\downarrow\uparrow}e^{-ik(n+1)a} \\
        e^{ik(n+1)a} + r_{\downarrow\downarrow}e^{-ik(n+1)a}
    \end{array}\right)$ and \\
$\mbox{\boldmath $B_n$}=\left(\begin{array}{cc}
    \tau_{\downarrow\uparrow}e^{ikna} \\
    \tau_{\downarrow\downarrow}e^{ikna}
    \end{array}\right)\,$,
\vskip 0.2cm
\noindent
respectively. $\tau_{\downarrow\uparrow}$ ($\tau_{\downarrow\downarrow}$) and
$r_{\downarrow\uparrow}$ ($r_{\downarrow\downarrow}$) are the transmission
and reflection amplitudes for down spin transmitted and reflected as up (down) spin,
respectively.

Using the same prescription as stated for the case of up spin incidence,
we calculate the transmission probabilities and bond
current densities for the down spin incidence as follows
\begin{eqnarray}
T_{\downarrow\sigma} & = & |\tau_{\downarrow\sigma}|^2
\label{tran2}
\end{eqnarray}
and
\begin{eqnarray}
J_{i\rightarrow i+1,\downarrow\sigma} &=& \frac{\sqrt{t^2+\alpha^2}\mbox{Im}
\left[C_{i,\downarrow\sigma}^*C_{i+1,\downarrow\sigma} \right]}{(1/2)t_0\sin(ka)},
\sigma\rightarrow\uparrow,\downarrow,\nonumber \\
\label{j2}
\end{eqnarray}
respectively.
\vskip 0.3cm
From the bond current density, we finally calculate the circular current density flowing
within the ring as:
\begin{eqnarray}
J_{\sigma,\sigma'} & = & \frac{1}{N} \sum_i J_{i\rightarrow i+1,\sigma \sigma'},
~~\sigma,\sigma'\rightarrow\uparrow,\downarrow.
\label{j3}
\end{eqnarray}

\subsection{Circular Current}

The net circular current within the ring, for a particular bias voltage $V$ at
absolute zero temperature, can be evaluated from the relation~
\begin{equation}
I_{\sigma\sigma'}(V) = \int\limits_{E_F-\frac{eV}{2}}^{E_F+
\frac{eV}{2}}J_{\sigma\sigma'}(E) \, dE, ~~~~\sigma,\sigma'\rightarrow\uparrow,\downarrow.
\label{j4}
\end{equation}
$E_F$ is the equilibrium Fermi energy. The net up and down-spin currents are defined as:
\begin{eqnarray}
I_{\uparrow} & = & I_{\uparrow\uparrow} + I_{\downarrow\uparrow} \nonumber, \\
I_{\downarrow} & = & I_{\uparrow\downarrow} + I_{\downarrow\downarrow},
\label{cur1}
\end{eqnarray}
respectively~\cite{cir11h}.
Using $I_{\uparrow}$ and $I_{\downarrow}$, we define the net charge and spin currents as
\begin{eqnarray}
I_C & = & I_{\uparrow} + I_{\downarrow}, \nonumber \\
I_S & = & I_{\uparrow} - I_{\downarrow},
\label{cur2}
\end{eqnarray}
respectively~\cite{cir11h}.

\section{Results and discussions}

There are a few parameters that are kept constant throughout the paper. The onsite potentials are
chosen to be zero, i.e., $\epsilon_0 = \epsilon_n = 0~\forall~n$.
The nearest-neighbor hopping integrals are taken as: $t_0 = 2\,$eV, $t = 1\,$eV, and $t_S = t_D = 0.5\,$eV. The
source is always connected to the first site of the ring, which is $N_S = 1$. We consider the lattice
spacing $a$ as $1\AA$. Current moving at the counter-clockwise direction in any segment of the ring
is considered to be positive.

\subsection{Without Spin Orbit Interaction}

\begin{figure}[ht]
{\centering \resizebox*{8cm}{2.7cm}{\includegraphics{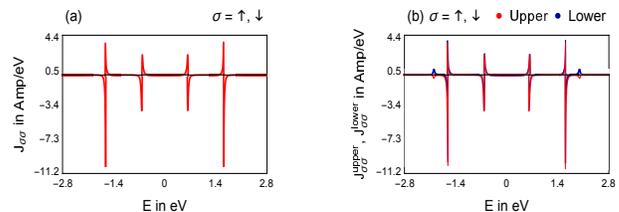}}\par}
\caption{(Color online). (a) Current density $J_{\sigma\sigma}$ as a function of
energy $E$. (b) Current densities flowing at the upper and lower arms of the ring.}
\label{fig1}
\end{figure}

First, we try to understand the basic features of the current density without any spin-orbit
interaction. When the ring is
symmetrically connected to the source and drain, the net circular current
becomes zero. Therefore we concentrate on asymmetric ring-to-lead configuration
(Fig.~\ref{fig1}).
The ring has $10$ atomic sites. The drain is
connected at the 7-th site of the ring. In Fig.~\ref{fig1}(a)
we plot the $J_{\sigma\sigma}$ ($\sigma=\uparrow$ or $\downarrow$) with energy $E$.
Both the $J_{\uparrow\uparrow}$ and $J_{\downarrow\downarrow}$ are the same, as
no spin scattering interaction is present in the system. For
the same reason, the spin current densities corresponding to the spin flipping
process ($J_{\uparrow\downarrow}$ and $J_{\downarrow\uparrow}$) are also
zero here. The energies associated with the picks and the dips in the spectra
correspond to the energy eigenvalues of the ring. For our present choice
of parameter values, the eigenvalues of the ring Hamiltonian
(written in Eq.~(\ref{hamil})) without any spin-orbit interaction ($\alpha=0$) are:
$-2$, $-2$, $-1.62$, $-1.62$, $-1.62$, $-1.62$, $-0.62$, $-0.62$, $-0.62$,
$-0.62$, $0.62$, $0.62$, $0.62$, $0.62$, $1.62$, $1.62$, $1.62$, $1.62$, $2.0$, $2.0\,$eV.
Along with the two fold spin degeneracy of each energy level, the system has
another doubly degenerate energy levels due to the periodic boundary condition
$N+1\equiv 1$ which
leads to the energy dispersion as $E = 2 t \mbox{Cos}(ka)$ where $k = 2 \pi m/N a$.
The integer $m$ runs between $N/2 \leq m <N/2$. Therefore for $k$ and $-k$, the system
has the same energy except at $m=0$ for odd $N$ and $m= -N/2, 0$ for even $N$.
For example, in our present setup, $N =10$. Therefore the degeneracies appear when
$m = \pm 4, \pm 3, \pm 2$, and $\pm 1$.
Whereas the energy levels corresponding to $m = -5$, that is $E = -2\,$eV and
$m = 0$ with $E = -2\,$eV, remain non-degenerate. The doubly degenerate orbitals are
characterized by their orbital angular momentum, representing Bloch waves traveling
clockwise or counter-clockwise along the ring. The circular current appears within the
ring when this degeneracy is lifted due to the ring-to-leads coupling. The symmetric
ring-to-lead connection does not split the degeneracies hence a net
circular current (specifically charge circular current) does not appear. Whereas, for
an asymmetric connection, there is a net current within the ring.

The current densities flowing through the upper
and lower arms ($J_{\sigma\sigma}^{upper}$ and $J_{\sigma\sigma}^{lower}$, respectively)
are plotted in Fig.~\ref{fig1}(b). As we can see, $J_{\sigma\sigma}^{upper}$
\begin{figure*}[ht]
{\centering \resizebox*{11cm}{7.8cm}{\includegraphics{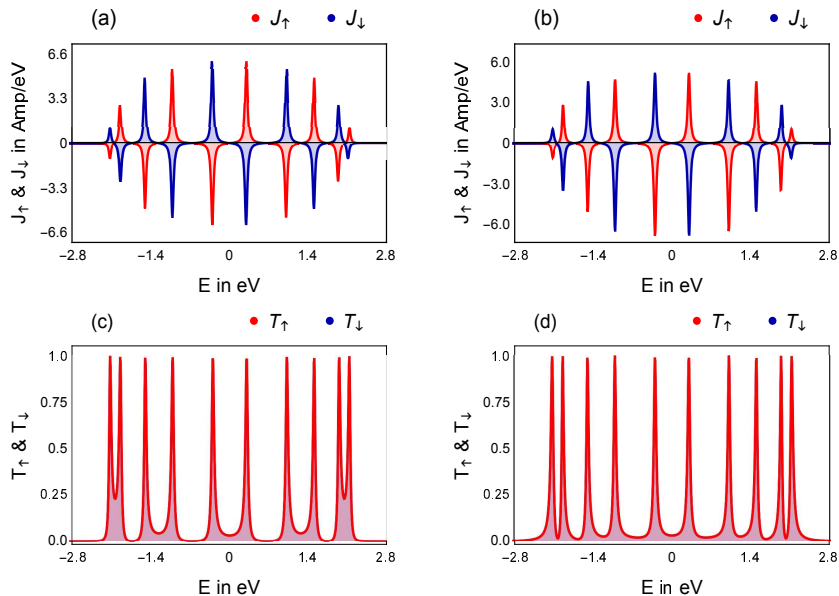}}\par}
\caption{(Color online). (a) - (b) Up and down spin current densities and (c) - (d)
up and down spin transmission probabilities with energy for $\alpha = 0.4\,$eV.
The left column ((a) and (c)) represent symmetric ring-to-lead connection
whereas right column i.e. (b) and (d) are simulated
for most asymmetric configuration (i.e., $N_D = 10$). The other parameters are same as
Fig.~\ref{fig1}.}
\label{fig3}
\end{figure*}
is opposite in sign to the $J_{\sigma\sigma}^{lower}$ at non-degenerate energy levels.
But they flow in the same direction at degenerate energy levels.
The slight splitting at these degenerate levels are caused by the coupling of
the ring with the electrodes.
The current density in terms of $J^{\mbox{\tiny upper}}_{\sigma}$ and
$J^{\mbox{\tiny lower}}_{\sigma}$ can be written as
\begin{equation}
J_{\sigma} = f^{\mbox{\tiny upper}}J^{\mbox{\tiny upper}}_{\sigma} +
f^{\mbox{\tiny lower}}J^{\mbox{\tiny lower}}_{\sigma}.
\label{eqR1}
\end{equation}
$f^{\mbox{\tiny upper}} = (N_D - 1)/N$ and $f^{\mbox{\tiny lower}} = (N - N_D + 1)/N$
are the weight factors for the upper and lower arms, respectively.
As across $E = \pm 2\,$eV,
the current flows in the two arms of the ring in opposite directions, with almost equal magnitude,
vanishingly small current densities are obtained.
Whereas at the degenerate energies (neglecting spin degeneracy),
the contributions from both of the arms are additive, hence a net circular current
density is obtained.


\subsection{With Spin Orbit Interaction}

The Rashba spin-orbit interaction causes a momentum-dependent spin splitting of electronic bands. But
with non-zero SOI, the quantum ring still has at least one more eigenstate with the same energy
according to the Kramers degeneracy theorem as our spin-half system preserves time-reversal symmetry.
Circular current (as well as non-zero transmission probability) appears for corresponding energy
eigenvalues similar to the previous situation
(for Rashba spin-orbit interaction strength $\alpha = 0.4\,$eV and with our present choices of
parameters, these energies are: $-2.12$, $-2.12$, $-1.94$,
$-1.94$, $-1.5$, $-1.5$, $-1.01$, $-1.01$, $-0.3$, $-0.3$,
$0.3$, $0.3$, $1.01$, $1.01$, $1.5$, $1.5$, $1.94$, $1.94$, $2.12$, and $2.12\,$eV.)
The effect of spin-orbit interaction on current density as well as on the transmission
spectra is studied in Fig.~\ref{fig3} for symmetric and asymmetric connections.

As for the two terminal SOI device, magnetic field (to break the time reversal symmetry) is a
\begin{figure}[ht]
{\centering \resizebox*{7.5cm}{2.5cm}{\includegraphics{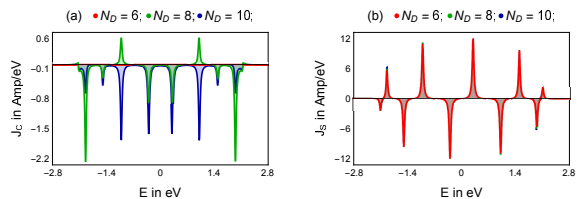}}\par}
\caption{(Color online). Charge ($J_C$) and spin ($J_S$) current densites with
energy $E$ for $N=10$ and $\alpha = 0.4\,$eV. In each case we calculate the current
for three different ring-to-lead configurations. The red curve corresponds to
$N_D = 6$, whereas for green and blue curves we choose $N_D = 8$ and 10,
respectively.}
\label{fig4}
\end{figure}
key ingredient to produce an net spin-polarized current at outgoing terminal, in the transmission
probability we do not observe any spin-separation for symmetric (Fig.~\ref{fig3}(c)) as
well as asymmetric (Fig.~\ref{fig3}(d)) cases. But within the conductor a net spin current
appears for both the cases (Fig.~\ref{fig3}(a) - (b)).
In fact for the symmetric case (Fig.~\ref{fig3}(a)), we have
\begin{eqnarray}
J_{\uparrow}(\pm E) & = & -J_{\downarrow}(\pm E).\nonumber \\
\label{purespin}
\end{eqnarray}
Therefore, in this situation,
throughout the energy window, the net charge current density
\begin{eqnarray}
J_C(E) & = & J_{\uparrow}(E) - J_{\downarrow}(E) = 0.
\label{chargeS}
\end{eqnarray}
But net spin current density,
\begin{eqnarray}
J_S(E) & = & J_{\uparrow}(E) - J_{\downarrow}(E) \nonumber \\
& = & 2 J_{\uparrow}(E) = -2 J_{\downarrow}(E).
\label{spinS}
\end{eqnarray}
\noindent
Hence a pure spin current appears (charge current is zero).
Apart from equality relations stated in Eq.~(\ref{purespin}),
for symmetric connection, we also have
\begin{eqnarray}
J_{\uparrow}(E) & = & -J_{\uparrow}(-E), \nonumber \\
J_{\downarrow}(E) & = & -J_{\downarrow}(-E). 
\label{purespin2}
\end{eqnarray}
\noindent
Equation~(\ref{purespin2}) implies,
\begin{eqnarray}
J_{S}(E) & = & J_{\uparrow}(E) - J_{\downarrow}(E) \nonumber \\
& = & J_{\downarrow}(-E) - J_{\uparrow}(-E) \nonumber \\
& = & - J_S(-E).
\label{r2}
\end{eqnarray}
\noindent
As the net spin current $I_S$ at a voltage $V$
is given by the area under the $J_S - E$ (Eq.~(\ref{j4})), therefore
net $I_S$ is zero under the condition $E_F = 0$. Therefore to get pure spin current,
we need to set Fermi energy $E_F$ other than zero.

For the asymmetric connection (Fig.~\ref{fig3}(b)), when there is a splitting in the degeneracy,
we have a net charge as well as spin circular currents. In this condition,
\begin{figure}[ht]
{\centering \resizebox*{6.5cm}{4cm}{\includegraphics{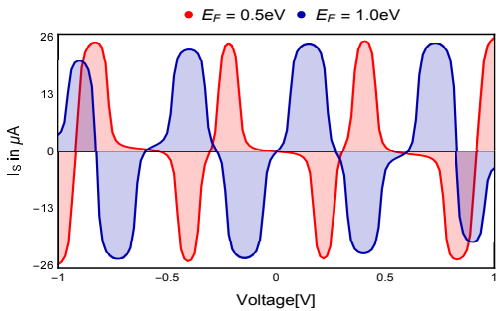}}\par}
\caption{(Color online). Superposition of net spin currents $I_S$s with bias voltage $V$
for a symmetric (red curve) and most asymmetric (blue curve) connections. The results
are computed for a 40 site ring for two different Fermi-energies ($E_F$). The strength
of spin-orbit interaction is $0.2\,$eV.}
\label{fig5}
\end{figure}
the up and down components of the circular current density are opposite to each other
but they are not equal, that is:
\begin{eqnarray}
J_{\uparrow}(\pm E) & \neq & - J_{\downarrow}(\pm E).\nonumber \\
\label{chargespin1}
\end{eqnarray}
\noindent
But similar to the symmetric connection condition, for asymmetric connection we still have,
\begin{eqnarray}
J_{\uparrow}(E) & = & J_{\downarrow}(-E),\nonumber \\
J_{\downarrow}(E) & = & J_{\uparrow}(-E).
\label{chargespin2}
\end{eqnarray}
\noindent
Therefore, under asymmetric connection,
\begin{eqnarray}
J_C(E) & = & J_{\uparrow}(E) + J_{\downarrow}(E) \nonumber \\
& = & J_{\downarrow}(-E) + J_{\uparrow}(-E) \nonumber \\
& = & J_C(-E).
\label{chargeAS}
\end{eqnarray}
\noindent
But for the net spin current we have,
\begin{eqnarray}
J_S(E) & = & J_{\uparrow}(E) - J_{\downarrow}(E) \nonumber \\
& = & J_{\downarrow}(-E) - J_{\uparrow}(-E) \nonumber \\
& = & - J_S(-E).
\label{spinAS}
\end{eqnarray}
\noindent
Therefore the spin $I_S$ current vanishes for $E_F = 0$ similar to the symmetric
connection situation. But as the charge current density is symmetric around $E = 0$,
we can get a pure charge current setting the Fermi energy at 0.

The total charge current density $J_C$ is plotted in Fig.~\ref{fig4}(a) for
three different ring-to-leads configurations. In Fig.~\ref{fig4}(b), we
calculate the spin current density for the same, though as they are almost similar.
There a are few basic differences between $J_C$ and $J_S$ that we can see in
Fig.~\ref{fig4}. (i) In each case a total of 10 peaks and dips are
visible there but for spin current density the subsequent energy levels
carry currents in opposite directions, whereas no such
sequence is seen for charge current density. (ii) $J_C$ is symmetric around
$E=0$, whereas $J_S$ is anti-symmetric. In other words $J_C(E) = J_C(-E)$ and
$J_S(E) = - J_S(-E)$, as we have already found in Eq.~(\ref{r2}), Eq.~(\ref{chargeAS})
and \ref{spinAS}. (iii) The charge current density is very much sensitive to the
ring-lead connection positions, but $J_S$ is quite independent of that.

\subsection{Spin current}

Now we calculate the total pure spin current.
\begin{figure}[ht]
{\centering \resizebox*{7.5cm}{2.5cm}{\includegraphics{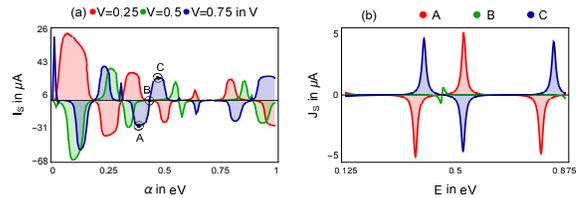}}\par}
\caption{(Color online).(a) Variation of pure spin current $I_S$ with Rashba spin-orbit
interaction $\alpha$ in a symmetric $30$-size junction at three different
voltages. (b) Spin current densities at three different values of $\alpha$,
those are marked by encircled dots in (a).}
\label{fig6}
\end{figure}
In Fig.~\ref{fig5} we plot the pure spin current $I_S$ with voltage $V$ setting Fermi energy at
$E_F = 0.5\,$eV (red) and $E_F = 1\,$eV (blue). The ring has
40 atomic sites and is symmetrically connected to the drain at $N_D =21$.
$I_S$ shows oscillation with voltage $V$, for both the choices of the Fermi
energies. For a very small voltages
around zero, the current is vanishingly small as no resonant energy level appears
within the window. Current becomes finite when anyone of such energy levels lies
within the voltage window. As we further increase the voltage, more and more resonant
energy levels appear within the window. Depending on their contributions to the net
current, the circular current becomes positive or negative, or zero. As the spin
circular current density remains almost unchanged with the connection position
of the drains, the current-voltage spectra are almost the same, for the other
ring-to-lead configurations with fixed $N$.

In order to see the dependence of pure spin current on spin-orbit
interaction we present its variations as a function of $\alpha$ for some 
typical values of bias voltage $V$ in Fig.~\ref{fig6}(a). 
$I_S$ has an oscillatory behavior with $\alpha$ and its sign alternately
changes from positive to negative for a wide window of $\alpha$. Therefore
we can control the $I_S$ by spin-orbit interaction without disturbing any
physical parameters of the system
and can be utilized in designing effective spin-based quantum devices.
To explain this large variation of the spin current with SOI, we choose
three distinct points A, B, and C from $I_S-\alpha$ curve of Fig.~\ref{fig6}(a),
represented by encircled dots, and present the current densities for the corresponding
values of $\alpha$ in Fig.\ref{fig6}(b). The results are shown for a specific energy
window $(0.125 \leq E \leq 0.875)$ associated with the voltage $V = 0.75\,$V and
Fermi-energy $E_F = 0.5\,$eV. For the $\alpha$ value associated with point A,
two current density peaks appear at negative energy while there is only one positive
energy peak (shown by the red color curve in Fig.\ref{fig6}(b)),  which results in a net
negative circular current. The scenario gets reversed at the $\alpha$ value
associated with point C (shown by the blue color in Fig.\ref{fig6}(b)). Therefore
a net positive current flows in the ring. At the $\alpha$ value associated with point
B, the current densities obtained for both positive and negative energies are
closely equal (shown by the green color curve in Fig.\ref{fig6}(b)). Therefore
vanishing spin current appears in this case. 

\section{Conclusion}

In summary, we have discussed the effects of Rashba spin-orbit interaction on the
circular current, which appears within a conductor having loop geometry.
We have discussed the origin of the circular currents from the degeneracy point of view.
We have found that, in a symmetric junction, the charge circular current is
always zero, but in this case, we have got non-zero pure spin circular current
(i.e., minimization of the Joule heating) setting the Fermi-energy at any value other than
zero. The system has
double degeneracy which can be characterized by its orbital angular momentum. On the other hand,
in an asymmetric junction, we have got a pure charge current, setting Fermi-energy
at zero. Due to ring-to-electrodes coupling, the degeneracy is removed here.
Finally, we have shown a way to regulate the pure spin current by changing the
strength of spin-orbit interaction. Our results will serve to design the new generation
spintronic devices where only spin will carry the information.

\section{Acknowledgements}

The author  acknowledges the financial support by the Postdoctoral Fellowship of
Indian Institute of Science Education and Research, Pune, India and the Japan
Society for the Promotion of Science Postdoctoral Fellowship
for Research in Japan (JSPS, ID No. P21022). The Author is thankful to Bijay
Kumar Agarwalla and Santanu K. Maiti for numerous useful discussions.

\end{document}